\documentclass[%
 reprint,
 amsmath,amssymb,
 aps 
]{revtex4-2}

\usepackage{graphicx}
\usepackage{dcolumn}
\usepackage{bm}

\begin{document}

\preprint{APS/123-QED}

\title{An Invertible All-optical Logic Gate on Chip}
\author{Zhan Li $^{1,2}$, Jiayang Chen $^{1,2}$, Zhaohui Ma $^{1,2}$, 
Chao Tang $^{1,2}$, Yong Meng Sua $^{1,2}$, Yu-Ping Huang $^{1,2,3,\dag}$}
\affiliation{$^1$Physics Department, Stevens Institute of Technology, Hoboken, New Jersey, 07030, United States \\
$^2$Center for Quantum Science and Engineering, Stevens Institute of Technology, Hoboken, New Jersey, 07030, United States\\ 
$^3$Quantum Computing Inc, Hoboken, NJ, 07030, United States \\
$^\dag$yuping.huang@stevens.edu}


\begin{abstract} 
We demonstrate an invertible all-optical gate on chip, with the roles of control and signal switchable by slightly adjusting their relative arrival time at the gate. It is based on quantum Zeno blockade driven by sum-frequency generation in a periodic-poled lithium niobate microring resonator. For two nearly-identical nanosecond pulses, the later arriving pulse is modulated by the earlier arriving one, resulting in 2.4 and 3.9 power extinction between the two, respectively, when their peak power is 1 mW and 2 mW. Our results, while to be improved and enriched, herald a new paradigm of logical gates and circuits for exotic applications.    
\end{abstract}

\maketitle

\section{Introduction}
In transistors and switches, either electronic or optical, the roles of control and signal are fixed \cite{fortunato2012oxide, el2008optical}. They often enter the circuits through distinct paths, with the control to change the physical state of the device, and the signal experiencing the change to implement a logical operation: the signal output altered by the presence of the control. Such operations can be (i) fully classical, where both the control and signal are describable classically, such as semiconductor XOR and AND gates; (ii) semi-nonclassical, where either needs to adopt a quantum description, like quantum switches \cite{huang2012quantum,oza2013entanglement}; and (iii) fully quantum mechanical, where both must adopt the quantum description, such as quantum CNOT gates \cite{torma2014strong}.   

Yet quantum physics can pave another avenue to logic: the signal can be altered by the control without the latter having to change the physical state of the device. Instead, just the mere potential of such occurrence can cause the signal be modulated or switched, an interesting phenomenon related to ``interaction-free measurement''  \cite{elitzur1993quantum,kwiat1995interaction}. The underlying physics is the quantum Zeno effect, based on which counter-intuitive quantum processing can be implemented, including counterfactual quantum computing \cite{hosten2006counterfactual}, counterfactual quantum communications \cite{salih2013protocol}, antibunched photon-pair emission \cite{huang2012antibunched}, and quantum Zeno blockade (QZB) \cite{huang2011interaction,jacobs2009all,huang2010interactionf}. 

Here, we study another interesting feature of the Zeno-based logical gates. We show that, in addition to the interaction-free features, QZB also allows the roles of the control and signal to be swapped by just slightly adjusting their relative arrival at the gate. This is distinct to all existing logical gates, where their roles are fixed. As such, one may invert the logic using a variable delay line, and complex functions can be realized by cascading multiple QZB gates. This opens a door to new computing circuitry and architectures, offering prospects of high efficiency, low power consumption, and unprecedented capabilities.   

Our demonstration uses a nonlinear cavity circuit on thin-film lithium niobate (TFLN), a strong contender of next-gen optical processing \cite{zhu2021integrated,lin2020advances}. Among many advantageous circuits TFLN supports, periodic-poled TFLN microrings can induce strong photon-photon interaction, \cite{chen2019ultra, lu2020toward}, with single-photon coupling strength reaching 8.2 MHz \cite{chen2021photon}. This allows efficient all-optical modulation with record low power consumption \cite{li2024parametric}. In our experiment, two optical pulses of different wavelengths but nearly identical shapes and peak power are quasi-phase matched for sum frequency generation (SFG) in the cavity. One pulse arrives slightly ahead of the other, and couples first into the cavity. When the second arrives, it ``sees'' an already-occupied cavity with the potential for SFG. The SFG would cause the signal loss to a sum-frequency mode, thereby implementing effective measurement to cause QZB. With strong SFG, the signal will thus not enter the cavity (in the asymptotic limit), so that the switch or modulation is realized in the ``interaction-free'' manner \cite{huang2011interaction,jacobs2009all,huang2010interactionf,huang2012antibunched, mccusker2013experimental,guo2018all,chen2017observation}. This effect can also be intuitively understood as this: when the SFG-induced loss is so high, the cavity becomes effectively under-coupled, so that the second pulse will mostly not enter the cavity. As such, the SFG nearly does not happen, so that the first pulse will not be affected by the second. Therefore, the first pulse always acts as the control and the second as the signal. Interestingly, their roles are only defined by which arriving at the cavity first. This realizes an invertible all-optical logical operation by adjusting the relative arrival time, which could be tens to hundreds of picoseconds.

\begin{figure*}[ht]
\centering
\includegraphics[width=0.8\linewidth]{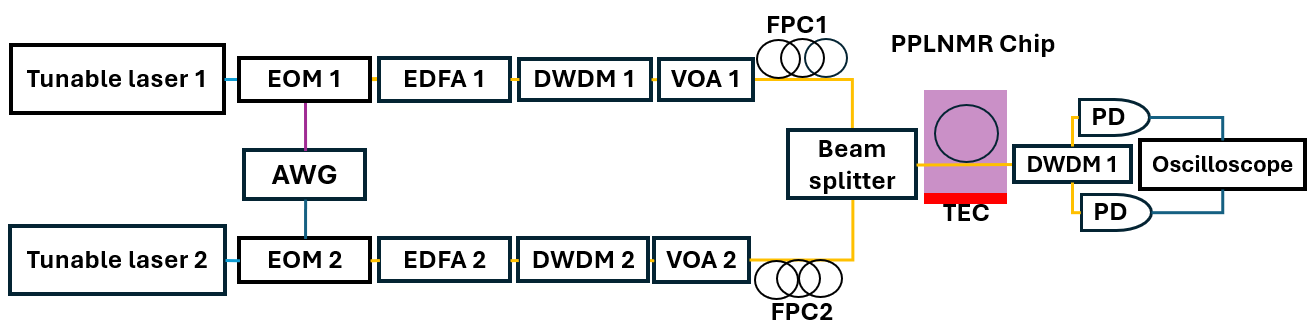}
\caption[Experiment schematic]{Experiment schematic. Orange lines indicate optical connections while blue lines are for electronic connections. The abbreviations in the figure represent as following: electro-optical Modulator (EOM); arbitrary waveform generator (AWG); Erbium-doped fiber amplifiers (EDFA); variable optical attenuators (VOA); thermo-electric cooler (TEC); dense wavelength-division multiplexing (DWDM, 200GHz), DWDM 1 central wavelength at 1545.35 nm and DWDM 2 at 1557.9 nm. Lensed fibers are used for coupling lights in and out chip. The fiber-chip coupling loss is measured 5 dB per facet.}
\label{fig:setup}
\end{figure*} 


\section{Experiments and simulation}
The periodically poled TFLN microring is fabricated on a Z-cut 600 nm TFLN, with 80 $\mu$m ring radius, 1700 nm top-width and 149 poling grating number. The detailed fabrication and parameter design can be found in our previous studies \cite{chen2019ultra,ma2020ultrabright,chen2021photon,ma2023highly}. To make the best use of the largest nonlinear tensor $d_{33}$, the quasi-phase matching conditions for achieving triple resonance are the fundamental quasi-transverse-magnetic (quasi-TM) modes. The photon-photon interaction strength $g$ is measured to be 8.2 MHz in angular frequency. The quantum conversion efficiency reaches 65\% when pump is 115 $\mu$W When it is "cold" (at low laser intensities), the cavity's resonances are all in the over-coupled regime. The triple resonances are identified at wavelengths of 1545.73 nm, 1558.15 nm, and 775.97 nm. In this study, we call pulse trains at 1545.73 nm and 1558.15 as Pulse 1 and Pulse 2, respectively. The cold-cavity total decay rate ($\kappa_{t,1}$) for 1545.73 nm stands at 9.4 GHz, with an external decay rate ($\kappa_{e,1}$) of 7.6 GHz. For the resonance at 1558.15 nm, the total decay rate ($\kappa_{t,2}$) is measured at 10.1 GHz, and the external decay rate ($\kappa_{e,2}$) is 8.5 GHz. For the sum-frequency resonance, the total decay rate ($\kappa_{t,f}$) is 14.1 GHz, and the external decay rate ($\kappa_{e,f}$) is 12.1 GHz.

The SFG in the nonlinear cavity can be described by the following simplified coupled mode equations \cite{bogaerts2012silicon}:
\begin{eqnarray}
& &\frac{dC_1}{dt}    =(i\delta_1-\frac{\kappa_{t,1}}{2})C_1-ig^*C^*_2C_{f}-i\sqrt{\kappa_{e,1}}F_1 \label{eq1} \\
& &\frac{dC_2}{dt}    =(i\delta_2-\frac{\kappa_{t,2}}{2})C_s-ig^*C^*_1C_{f}-i\sqrt{\kappa_{e,2}}F_2 \label{eq2}\\ 
& &\frac{dC_{f}}{dt} =(i\delta_{f}-\frac{\kappa_{t,f}}{2})C_{f}-igC_1C_2 \label{eq3}
\end{eqnarray}
Here, the subscripts $\sigma=1,2,f$ denote Pulse 1, Pulse 2, and the sum-frequency wave, respectively. $F_\sigma$ represents the amplitude of the input photon flux. $C_{\sigma}$ denotes the photon-number amplitudes in the cavity. $\delta_{\sigma}$ is the detuning of the laser frequency from the cavity's resonance, which ideally equals zero when phase matching conditions are perfectly met. The cavity's output, $O_{\sigma}$, is given by the interference of the light leaking out of the cavity and the incoming light, as $O_\sigma  =i\sqrt{\kappa_{e,\sigma}}C_\sigma+F_\sigma$. In our simulation,  the cold-cavity parameters are used along with the measured input pulses. The simulated outputs are calculated by solving the \eqref{eq1}, \eqref{eq2} and \eqref{eq3} in time domain with Matlab. The inputs of the simulation are the exact pulses in the experiments.

The experiment setup is depicted in Fig.(\ref{fig:setup}). Pulse 1 at 1545.73 nm and Pulse 2 at 1558.15 nm are generated from two tunable lasers (Santec TSL-550/570), each followed by an electro-optical modulator (EOM) to curve the pulses. The EOM's are driven by synchronized and delayed radio-frequency signal pulses generated by an arbitrary waveform generator (AWG). Each of the resulting light pulse trains is then amplified using an Erbium-doped fiber amplifier (SDH-EDFA-LA-023), with a 200GHz bandwidth dense wavelength-division multiplexing (DWDM) applied afterwards to get rid of the amplified spontaneous emission noise. Next, variable optical attenuators (Thorlabs VOA50-FC) are employed to adjust the optical power levels. A beamsplitter merges the optical pulses, which are then sent into the TFLN chip together via a lensed fiber. The coupling loss per facet is measured to be about 5.0 dB. The chip is precisely temperature-controlled and stabilized using a thermo-electric cooler (TEC), with an accuracy of $0.01^\circ C$. Post chip, another DWDM separates Pulse 1 and Pulse 2, and directs them to photodetectors (DET08CFC, 5 GHz) for measurement and an oscilloscope (HP Agilent 54750a, 20 GHz) for visualization. 

The pulses employed in the experiment are Gaussian-shaped, with a 2 ns full width half maximum (FWHM) and repetition rates at the megahertz level. This specific pulse configuration is chosen to mitigate the side effects associated with thermal-optic and photorefractive (PR) phenomena \cite{chen2020efficient,surya2021stable,shams2022reduced}. The relative arrival times of the pulses are finely adjusted by tuning the phase of the electrical signals on the AWG. To quantify the modulation effects, the measured output power of each pulse train is normalized against its measured peak power under the off-resonance condition, for a straightforward comparison of modulation efficiency across different experiment setups.

\begin{figure}[ht]
\centering
\includegraphics[width=1\linewidth]{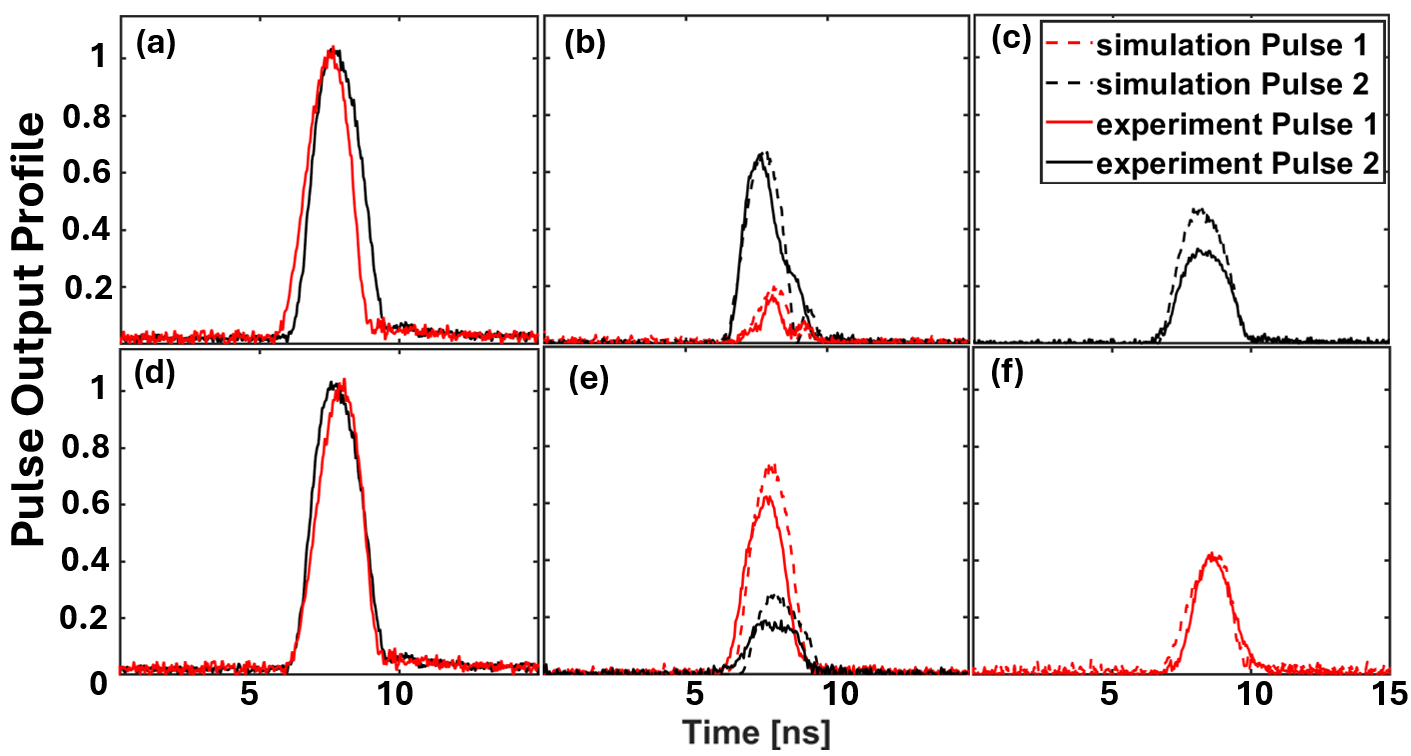}
\caption{Measured and simulated output pulses with input at 2 mW peak power. Figure (a) displays the off-resonance output when Pulse 1 (in red) precedes Pulse 2 (in black). Figure (b) depicts the outputs when both are on resonance. Figure (c) is Pulse 2 output when it is on resonance and Pulse 1 is absent. Figures (d)-(f) are for the same but with Pulse 2 preceding pulse 1.}
\label{fig:2mw}
\end{figure}

The experimental results when the peak power for both pulses are set at 2 mW is shown in Fig.~\ref{fig:2mw}. As studied previously, the modulation efficiency depends on their relative delay \cite{li2024parametric}. In this experiment, Pulse 1 (with a repetition rate of 1 MHz) is introduced into the cavity approximately 300 ps before Pulse 2 (with 2 MHz repetition rate). This results in Pulse 2 restoring its transmission through the cavity to 66.2\% of the off-resonance case due to QZB, as illustrated in Fig.~\ref{fig:2mw}(b). Meanwhile, the transmission of Pulse 1 is 16.8\% of that in Fig.~\ref{fig:2mw}(a), which is slightly reduced as compared to the on-resonance case in Fig.~\ref{fig:2mw}(f). This is because QZB is not strong enough to reject all Pulse 2, so that some SFG still occurs in the cavity to cause more loss to Pulse 1. While Pulse 1 and 2 have the same peak power, at the output the ratio between the two is 1:3.9. 

\begin{figure}[ht]
\centering
\includegraphics[width=1\linewidth]{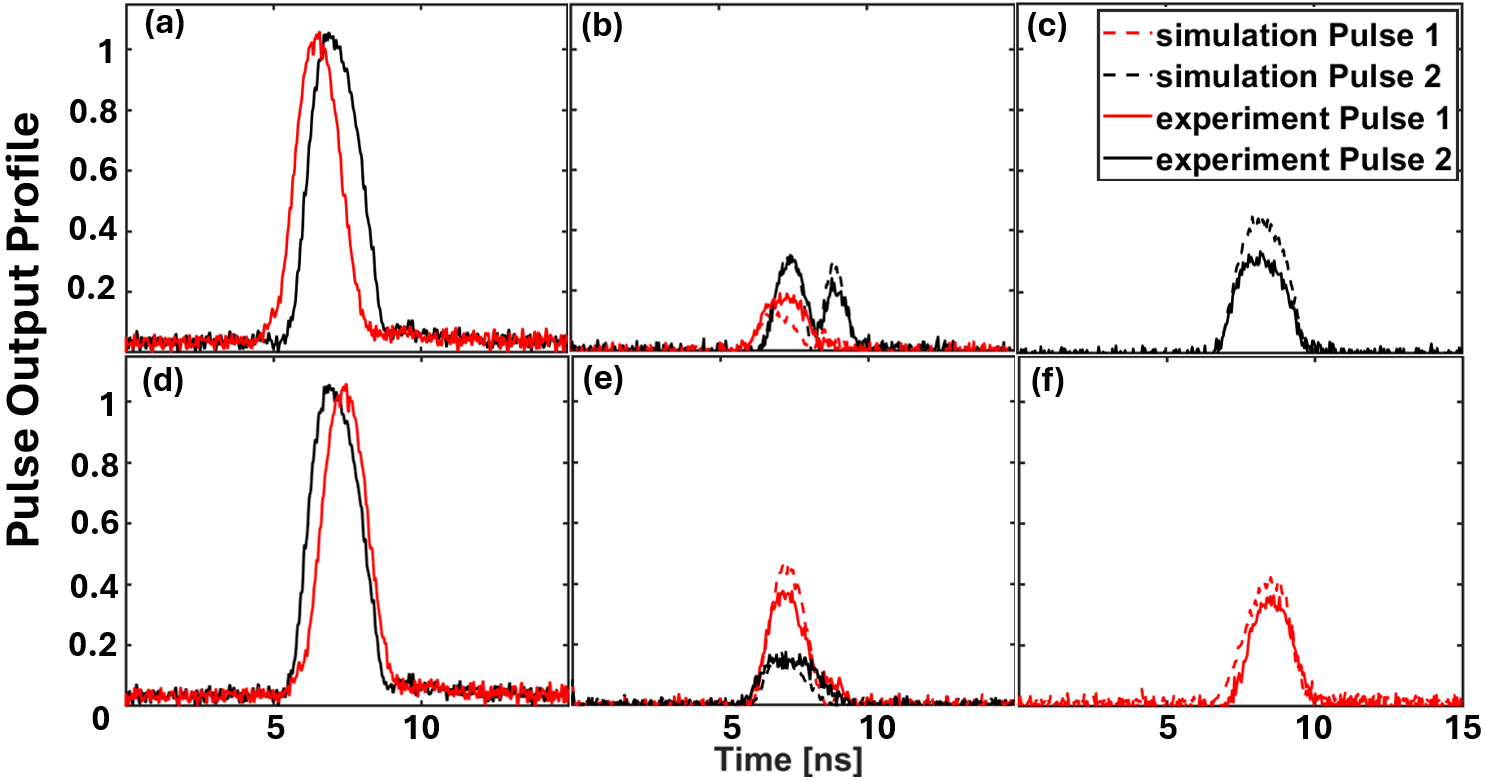}
\caption{Measured and simulated results similar to Fig.~\ref{fig:2mw} but with 1 mW peak power for both pulse trains.}
\label{fig:1mw}
\end{figure}

Subsequently, when Pulse 1 arrives roughly 300 ps after Pulse 2, their roles are inverted with a contrasting behavior: the transmission of Pulse 1 is restored to 62.3\%, whereas that of Pulse 2 decreases to 18.2\%; see Fig.~\ref{fig:2mw}(e). The power ratio is now 3.4:1, as opposed to 1:3.9 in Fig.~\ref{fig:1mw}(b). Note that in Fig.~\ref{fig:2mw}(e), the experimental results for Pulse 2 fall below the simulation predictions. This discrepancy could be attributed to a weaker experiment cavity coupling strength for Pulse 2, as suggested by Fig.~\ref{fig:2mw}(c), resulting in diminished power within the cavity and, consequently, lower modulation efficiency.

To delve deeper into the modulation efficiency, the pulse power is scaled down to 1 mW and the optimal modulation efficiency is achieved with the arrival time difference of approximately 500 ps. Following the same experimental procedures, the outputs for both pulse trains at this reduced power level are shown in Fig.~\ref{fig:1mw}. When Pulse 1 precedes Pulse 2, the Pulse 1 output is at 18.9\% of the off-resonance case, while Pulse 2 is restored to 37.8\%, as illustrated in Fig.~\ref{fig:1mw}(b). Conversely, when the roles are reversed with Pulse 1 arriving after Pulse 2, the former's output is 37.8\% and the latter drops to 15.5\%, as depicted in Fig.~\ref{fig:1mw}(e). The power extinction ratios for these two cases are 1:2.0 and 2.4:1, respectively. This is because the SFG-induced loss for each pulse is linearly reduced by half as compared to the 2-mW case in Fig.~\ref{fig:2mw}, so that the QZB modulation effect is reduced, too. Note that in Fig.~\ref{fig:1mw}(b), a side peak appears at the end of Pulse 2. This is because in our experiment, Pulse 2 is 200 ps longer than Pulse 1, due to the limitation of our available AWG. Consequently, towards the end of Pulse 1, there isn't sufficient power to induce QZB for Pulse 2. Rather, the portion of Pulse 2 overlapping with the trailing tail of Pulse 1 would experience higher loss due to SFG and become critically coupled. Further into the tail, the SFG is negligible, so that it returns to overcoupling, leading to the appearance of ringing. In Fig.~\ref{fig:1mw}(e), the discrepancy between the simulated and measured modulation efficiencies can once again be attributed to a weaker coupling strength for Pulse 2.

To counteract undesirable effects, including thermal-optic and photorefractive (PR) phenomena, the phase matching condition throughout the experiment is finely adjusted by tuning the lasers and the TEC. These side effects, which typically manifest at kilohertz-level response rates, occur approximately an order of magnitude slower than the repetition rate of the pulses. So, every other pulse in the 2-MHz pulse trains does not overlap with the other pulse train and serves as a monitoring mechanism to ensures the pulses remaining on-resonance and side effects offset. Figure \ref{fig:2mw}(c, f) and \ref{fig:1mw}(c, f) show such monitoring pulses.  

\section{Discussion and Conclusion}

This study demonstrates an invertible optical logical circuit on chip, where the roles of control and signal are swapped by changing the relative delay between two pulses. Strong QZB effect and all-optical modulation are observed with 10 pico-joule pulses. Our results point to a new paradigm of computing circuitry and architectures, where complex functions can be realized by cascading multiple QZB gates. 

For the next steps, the modulation efficiency can be further enhanced by using a microring with steeper and smoother sidewall for higher Q-factor and with a smaller effective mode volume for stronger photon interaction. Also, by incorporating a drop port along side the microring, a QZB switch can be realized where the optical paths of the  pulses can be swapped by adjusting their relative delay. 

Finally, the modulation for both pulses is ideally expected to be enhanced with higher input power. However, in the experiment, when the peak power is elevated to 3 mW, both pulses induce strong second harmonic generation (SHG) whose power is measured above 100 nW at the output. Together with the other stronger side effects, the cavity becomes too unstable to observe the modulation effect. Also, the simplified coupled mode equations in Eqs.(\ref{eq1}-\ref{eq3}) fail to predict the complex interactions in the cavity. The future studies will include suppressing SHG by using further detuned laser wavelengths or changing the cross-section geometry of the microring to enlarge the group-velocity mismatch between the two pulse trains.

\begin{acknowledgments}
The research was supported in part by the Office of Naval Research (Award No. N00014-21-1-2898). Device fabrication was performed in Nanofabrication Facility at Advanced Science Research Center (ASRC), City University of New York (CUNY). 

The authors declare no conflicts of interest.

Data underlying the results presented in this paper are not publicly available at this time but may be obtained from the authors upon reasonable request.
\end{acknowledgments}


\bibliography{sample}


\end{document}